# Deduction of Lorentz Transformation from the existence of absolute rest. Deduction of the speed of light in any frame of reference.

Rodrigo de Abreu
Centro de Electrodinâmica e Departamento de Física do IST

Abstract

We begin by admitting the following:

There is a frame of reference where the speed of light is the same in any direction – that speed is $c$.
The <u>average</u> speed of light on a two-way journey is $c$ in every frame of reference.

From this two premises we obtain an expression for the speed of light which implies the speed of light depends on the motion of the frame of reference. Also from this two premises solely we deduct Lorentz Transformation.

Introduction

Experience suggests that, when a ray of light is emitted towards a mirror and then reflected back to its point of origin, its average speed (the total distance divided by the total time) for the two-way journey is $c$ in every frame of reference. We'll assume this is in fact so. Also we'll assume the existence of a frame of reference where the speed of light has the value $c$ in every direction – we will assume light always moves with speed $c$ on this frame independently of the state of motion of the emitting body[1].

Obviously if the speed of light is $c$ in every direction in a given frame of reference, then the two-way average speed will also be $c$ in that frame. Yet, the two-way average speed being $c$ in a given frame, doesn't imply light travels with speed $c$ on both parts of the journey in that frame. So, while we assume the two-way average speed of light is $c$ in every frame, only for one frame do we assume the true speed of light is $c$ – we call this the resting frame.

The speed of light being $c$ in every direction in the resting frame means that if we simultaneously emit two rays of light in opposite directions they will travel the same distance in the same time in this frame. So, if we place two clocks at any two points at the same distance from the origin of the rays of light and if each ray of light is emitted towards a clock, then we can be sure the rays will reach the clocks simultaneously. In other words, we can safely use Einstein's method of synchronization [1] to synchronize all the clocks fixed on this frame of reference. On the other hand, for all frames except the resting one, we're not sure what the speed of light on a one-way journey is.

---

[1] Notice that while the speed of light may be independent of the state of motion of the emitting body, its value doesn't have to be $c$ in every frame of reference. We assume the speed of light must have constant value for every frame, but that value may change from one frame to other.

Therefore by using Einstein's method of synchronization, we <u>can't</u> be sure to synchronize the clocks of any frame except the resting one.

We'll next show that our premises are not only sufficient in order to obtain Lorentz Transformation, but also that they are inconsistent with the stronger premise that the speed of light is the same in every frame of reference [2].

## Lorentz Transformation obtained from absolute motion

Suppose the existence of a frame of reference $S$ where the speed of light is $c$ in any direction. Consider also a second frame of reference $S'$ moving with velocity $v$ related to $S$ – for simplicity's sake $S'$ only moves along the $x$-axis. We impose $y=y'$ and $z=z'$ for every $t$ (since they are perpendicular to the motion of $S'$, the y'-axis and the z'-axis do not contract). We'll suppose the clocks of $S$ are already synchronized between each other by Einstein's method – we can assume this, since we know the speed of light in $S$ is $c$ in any direction. We'll suppose that when the clocks of $S$ mark the instant $t=0$, all the clocks of $S'$ fixed at $x'=0$ also mark the instant $t'=0$. We'll also assume that when $t=0$ the origin of $S'$ coincides with the origin of $S$.

Now, suppose that by the moment the origin of $S'$ passes on the origin of $S$ (at $t=0$), a ray of light is emitted from that point along the positive direction of the y'-axis. Since $S'$ is only moving along the $x$-axis and this ray of light moves along the y'-axis and because we admit the two-way speed is always $c$, then we can conclude by symmetry that the one-way speed of this particular ray of light is also $c$ in $S'$ [2]. Yet we admit the one-way speed of the ray of light is also $c$ in $S$, so we must conclude the clocks of $S'$ are running slower. When we relate the movement of the ray of light in both frames of reference, we obtain the following relations between $t$ and $t'$:
.

$$c^2 t^2 = c^2 t'^2 + v^2 t^2 \qquad (1)$$

Therefore, from (1) we have

$$t = \frac{t'}{\sqrt{1 - \frac{v^2}{c^2}}} \qquad (2)$$

---

[2] This conclusion can be briefly explained by the following: Suppose the ray of light was reflected by a mirror fixed anywhere on the y'-axis - the ray would return to the origin. Since the ray moves parallel to the y'-axis, if we look at that two-way journey from the point of view of $S$, we will conclude both parts of the journey have the same distance in $S$. This means both parts of the journey take the same time to be completed in $S$ since light always moves with speed $c$ in this frame. But if this is so, then while it might take longer to complete each part of the journey in $S'$ then in $S$, the time of both parts must also be the same in $S'$ and so must the speed of light. Since the average speed of the total journey is $c$, each one-way journey must also be $c$. Of course this conclusion is only true because the light is moving parallel to the y'-axis.

Suppose that, also at *t=0*, a second ray of light is emitted from the coinciding origins towards a mirror fixed on the *x'*-axis on the point (*x'=x'*, *y '=0*, *z'=0*). Since the speed of light is *c* in *S* and the velocity of *S'* is *v* in *S*, we have the following equations for the position of the ray of light (3) and the mirror (4) in the frame of reference *S*:

$$x = x_0 + vt \qquad (3)$$

$$x = ct \qquad (4)$$

From (3) and (4) we obtain the time $t_1$ required for the light to travel from the coinciding origins to the mirror:

$$t_1 - 0 = \Delta t_1 = \frac{x_0}{c - v} \qquad (5)$$

We can, from a similar analysis, determine the time $t_2$ needed for the light to travel back from the mirror to the origin of *S'*. That time is:

$$\Delta t_2 = \frac{x_0}{c + v} \qquad (6)$$

From (5) and (6) we have the total time of the two-way journey:

$$\Delta t_1 + \Delta t_2 = \frac{x_0}{c - v} + \frac{x_0}{c + v} \qquad (7)$$

Or

$$\Delta t_1 + \Delta t_2 = \frac{2 x_0}{c(1 - \frac{v^2}{c^2})} \qquad (8)$$

Suppose the first ray of light, emitted along the *y'*-axis, is also reflected by a mirror fixed on this axis at the same distance from the origin (*x'=0*, *y' =x'*, *z'=0*). Since the speed of this ray is *c*, if *t'* is the time the light takes to reach the mirror on the *y'*-axis, then *t'* is also the time for the light to return to the origin. Since the total distance of each two-way journey of the two rays of light is the same in *S'* and since the average two-way speed of both rays of light is *c*, then the two rays of light take the same time to complete the journey. We have the following relation:

$$\Delta t_1 + \Delta t_2 = \frac{2t'}{\sqrt{1 - \frac{v^2}{c^2}}} = \frac{2x_0}{c(1 - \frac{v^2}{c^2})} \qquad (9)$$

And we have also:

$$2x' = c2t' \quad (10)$$

or

$$x' = ct' \quad (11)$$

From (9) and (11) we obtain:

$$x_0 = x'\sqrt{1 - \frac{v^2}{c^2}} \quad (12)$$

From (12) we see that the distance of the mirror fixed on the $x'$-axis from the coinciding origins is larger if measured in $S'$ then it is in $S$. We conclude that the $x'$-axis is contracted.

Since we know the relation between the time of $S$ and $S'$ (2) we can immediately obtain the time the light takes to reach the mirror fixed on the $x'$-axis (13) and the time the light takes to return to the origin of $S'$ (14). From (2), (5) and (6) we have:

$$\Delta t'_1 = \frac{x_0}{(c-v)}\sqrt{1 - \frac{v^2}{c^2}} \quad (13)$$

and

$$\Delta t'_2 = \frac{x_0}{(c+v)}\sqrt{1 - \frac{v^2}{c^2}} \quad (14)$$

From (12), (13) and (14) we obtain the speed of light as it travels towards the mirror (15) and the speed of light as it travels in the opposite direction (16). We can see they are different:

$$v'_1 = \frac{x'}{\Delta t'_1} = \frac{c-v}{(1 - \frac{v^2}{c^2})} \quad (15)$$

and

$$v'_2 = \frac{x'}{\Delta t'_2} = \frac{c+v}{(1-\frac{v^2}{c^2})} \qquad (16)$$

Finally we obtain the Lorentz Transformation. From (1) and (12) we have:

$$x_0 = x - vt = x'\sqrt{1-\frac{v^2}{c^2}} \qquad (17)$$

or

$$x' = \frac{x-vt}{\sqrt{1-\frac{v^2}{c^2}}} \qquad (18)$$

Since

$$x' = ct' = \frac{x-vt}{\sqrt{1-\frac{v^2}{c^2}}} = \frac{ct-vt}{\sqrt{1-\frac{v^2}{c^2}}} \qquad (19)$$

We have:

$$t' = \frac{t - \frac{v}{c^2}x}{\sqrt{1-\frac{v^2}{c^2}}} \qquad (20)$$

**Conclusion**

We've shown that while the average speed of light on a two-way journey may be $c$ in every frame of reference, the speed of light changes accordingly to the absolute motion of the frame it is measured in. The speed of light can only be constant in a single frame of reference. This implies absolute rest. We've also shown that the existence of absolute rest, doesn't imply Lorentz transformation is not true. We've derived Lorentz transformation from the same premises that lead us to the conclusion that the speed of light is not a constant. This shows Lorentz Transformation can and must be physically interpreted in the context of absolute motion [2].